\documentclass[11pt,pre]{revtex4-1}
\usepackage{graphicx}
\usepackage{algpseudocode}
\usepackage{epstopdf}
\usepackage{subfigure}

\usepackage{soul}

\usepackage{amssymb}
\usepackage{amsthm}
\usepackage{amsmath}
\usepackage[subnum]{cases}
\usepackage{empheq}
\usepackage{color}
\usepackage{booktabs}
\usepackage[dvipsnames]{xcolor}

\newcommand*\diff{\mathrm{d}}
\algdef{SE}[DOWHILE]{Do}{doWhile}{\algorithmicdo}[1]{\algorithmicwhile\ #1}%

\begin{document}

\title{Collision model for fully-resolved simulations of flows laden with finite-size particles}

\author{Pedro Costa}\email{p.simoescosta@tudelft.nl}
\affiliation{Delft University of Technology, Laboratory for Aero and Hydrodynamics, Leeghwaterstraat 21, NL-2628 CA Delft, The Netherlands}
\author{Bendiks Jan Boersma}
\affiliation{Delft University of Technology, Energy Technology, Leeghwaterstraat 39, NL-2628 CA Delft, The Netherlands}
\author{Jerry Westerweel}
\affiliation{Delft University of Technology, Laboratory for Aero and Hydrodynamics, Leeghwaterstraat 21, NL-2628 CA Delft, The Netherlands}
\author{Wim-Paul Breugem}
\affiliation{Delft University of Technology, Laboratory for Aero and Hydrodynamics, Leeghwaterstraat 21, NL-2628 CA Delft, The Netherlands}

\date{\today}

\begin{abstract}
We present a collision model for particle-particle and particle-wall interactions in interface-resolved simulations of particle-laden flows. Three types of inter-particle interactions are taken into account: (1) long- and (2) short-range hydrodynamic interactions, and (3) solid-solid contact. Long-range interactions are incorporated through an efficient and second-order accurate immersed boundary method (IBM). Short-range interactions are also partly reproduced by the IBM. However, since the IBM uses a fixed-grid, a lubrication model is needed for an inter-particle gap width smaller than the grid spacing. The lubrication model is based on asymptotic expansions of analytical solutions for canonical lubrication interactions between spheres in the Stokes regime. Roughness effects are incorporated by making the lubrication correction independent of the gap width for gap widths smaller than $\sim 1\%$ of the particle radius. This correction is applied until the particles reach solid-solid contact. To model solid-solid contact we use a variant of a linear soft-sphere collision model capable of stretching the collision time. This choice is computationally attractive because it allows to reduce the number of time steps required for integrating the collision force accurately and is physically realistic, provided that the prescribed collision time is much smaller than the characteristic timescale of particle motion. We verified the numerical implementation of our collision model and validated it against several benchmark cases for immersed head-on particle-wall and particle-particle collisions, and oblique particle-wall collisions. The results show good agreement with experimental data.
\end{abstract}
\maketitle

\setlength{\abovedisplayskip}{5pt}
\setlength{\belowdisplayskip}{3pt}
\setlength{\abovedisplayshortskip}{5pt}
\setlength{\belowdisplayshortskip}{3pt}

\section{Introduction}
\label{sec:intro}
Flows laden with solid particles appear widely in both nature and industry. Examples are the transport of sediments in a river, the enhanced mixing due to the presence of particles in a fluidized bed reactor, and the flocculation/sedimentation processes in the treatment of drinking water. In many cases the flow is turbulent, the size of the particles is comparable to or larger than the Kolmogorov length-scale (i.e., the particles have a finite-size), and the volume fraction of particles may be considerably high such that inter-particle interactions are dynamically important \cite{Elghobashi1994}.

Studying flows laden with finite-size particles using interface-resolved direct numerical simulations (DNS) has recently become possible with the development of efficient numerical methods, such as immersed boundary methods (IBM) \cite{Balachandar2010}, together with the continuous increase in computing power. Such simulations provide detailed insight in the flow dynamics at the particle scale and beyond. The governing equations for the fluid phase and the particles are directly coupled with each other through the no-slip/no-penetration condition at the particles' surfaces (i.e., 2-way coupling) without the need of parameterizing the drag force between the phases. Also, long-range hydrodynamic interactions between particles (i.e., 4-way coupling) are naturally reproduced by these methods. However, when the particle volume fraction is high, additional models are required to account for short-range hydrodynamic solid-solid interactions (lubrication forces) and solid-solid contacts. Otherwise, the realism of the simulation may be compromised by a poor description of these interactions. For instance, by under-predicting lubrication-enhanced clustering of inertial particles, as observed for homogeneous isotropic turbulent flows \cite{Cate2004}. The challenge is to find a model able to reproduce short-range particle-particle and particle-wall interactions with the required realism and with little effect on the computational efficiency of the overall numerical algorithm.

We consider non-Brownian spherical particles, which are sufficiently large such that inter-surface forces as the Van der Waals force and the electrostatic double-layer force can be neglected \cite{Grasso2002}. Also, cohesive forces, which are relevant for wet granular media \cite{Groger2003}, are disregarded. We restrict the applicability of the model to cases where 4-way coupling is required, but where the solid volume fraction is not extremely high such that good description of the macroscopic outcome of the collision (i.e., relative velocity prior to and after contact) is sufficient to model the suspension dynamics.

Much work has been done in modeling of inter-particle (or particle-wall) collisions. Discrete element methods (DEM) have been successfully used to account for inter-particle collisions in simulations of gas-solid flows where hydrodynamic interactions between particles are negligible (e.g., \cite{Cundall1979}, \cite{Tsuji1993} and \cite{VanderHoef2004}). These collisions are often referred to as \textit{dry} collisions. More recently, some studies used these same collision models for reproducing particle-particle and particle-wall interactions in viscous liquids, commonly referred to as \textit{wet} collisions. In this case, fluid effects such as added mass, viscous dissipation and history forces become important \cite{Gondret2002}. 

The lubrication effects cannot be resolved by the overall numerical method (not without resorting to excessive grid refinement). This lack of spatial resolution can be circumvented by a closure model for lubrication interactions based on analytical solutions of these interactions in the Stokes regime (e.g., \cite{Ten2002}, \cite{Breugem2010}, \cite{Kempe2012} and \cite{Motta2013}).

Many studies used variants of the soft-sphere collision model of \citet{Cundall1979} to compute the contact forces, because of its computational advantages for simulating dense suspensions when compared to hard-sphere models \cite{Deen2007}. In the soft-sphere model, the normal force acting on the particle during a collision is computed from an equivalent linear spring-dashpot system in which the spring stiffness and dashpot coefficients are parameterized as function of the particle's elastic properties. A limitation of this approach when applied to particle-laden flows is that the collision must be resolved with a time step that can be several orders of magnitude smaller than the time step of the Navier-Stokes solver for the fluid flow. This happens because the characteristic time scale of solid-solid contact is in general orders of magnitude smaller than the smallest time scale present in the flow \cite{Hertz1882}, \cite{Legendre2006}. However, it is possible to artificially stretch the collision time to a multiple of the time step with which the particle motion is integrated. In some studies this was done by decreasing the value of the spring stiffness and checking resulting the collision time in a trial and error procedure  \cite{Feng2010}. This approach was avoided by others, who prescribed the desired collision time and computed the corresponding collision parameters by solving the equations of the harmonic oscillator (e.g., \cite{Breugem2010}, \cite{Kempe2012}, \cite{Motta2013}).

Experimental studies have shown that the fluid effects in the normal collision of a sphere onto a plane wall can be quantified by an \textit{effective} normal coefficient of restitution, $e_n$, defined as the ratio of the magnitudes of rebound and impact velocities. In particular, when experimental data of $e_n/e_{n,d}$ (where $e_{n,d}$ is defined in an analogous way as $e_n$ but for a collision in a dry system) are plotted against the particle impact Stokes number, $\mathrm{St}\equiv (1/9)\rho_p U_p D_p/\mu$ (where $\rho_p$, $U_p$, $D_p$ and $\mu$ are respectively the particle mass density, impact velocity, diameter and the fluid dynamic viscosity), the datasets for different fluids and particle types collapse in the same curve \cite{Legendre2006}. This suggests that $e_n$, $e_{n,d}$ and $\mathrm{St}$ are key parameters to describe a head-on wet collision. Hence, reproducing this scaling is an important test for any numerical method for resolving the flow conforming a particle combined with a collision model should pass. 
Several authors have been able to reproduce it with different methodologies for resolving the particle-fluid interface, such as tensorial penalty methods \cite{Motta2013}, Lagrange multiplier-based methods \cite{Ardekani2008} or IBM (\cite{Breugem2010}, \cite{Li2011}, \cite{Izard2014} and \cite{Kidanemariam2014}). However, this benchmark experiment relies in a definition of impact and rebound velocities, which vary significantly in these references \cite{Izard2014}. Hence, if one solely resorts to this simple benchmark for validating the head-on collision model without careful comparison with experimental data, it can happen that the definitions of the impact and rebound velocities determined from the numerical simulation are not consistent with the measured quantities.

The complexity of the problem increases when the collision is oblique. In this case, the relative motion between the contact surfaces has a tangential component. Two different kinds of motion can occur between the surfaces in contact: rolling and sliding. Rolling occurs when a point of contact has zero relative velocity with respect to the contact surface, otherwise sliding occurs. Moreover, when a particle flowing through a viscous liquid approaches a planar surface obliquely, it experiences not only lubrication forces due to the squeezing motion of the fluid through the gap, but also forces and torques due to relative translational and rotational shearing (see \cite{Dance2003} for a review). Finally, the frictional resistance of the contact surface in the presence of a viscous liquid can change abruptly due to piezoviscous effects when smooth particles collide obliquely \cite{Joseph2004}.

To the best of the our knowledge, \citet{Kempe2012} report the only collision model validated against experimental data of oblique particle-wall collisions in viscous liquids and against bouncing trajectories of particles colliding onto a planar surface in a viscous liquid. The latter benchmark validation is particularly interesting to reproduce because it does not rely on definitions of impact and rebound velocity. It therefore gives a finer indication of the success of the model to reproduce the canonical case of a particle-wall collision than reproducing data of $e_n/e_{n,d}$ vs $\mathrm{St}$. \citet{Kempe2012} computed the normal collision force from a non-linear spring-dashpot system. This was done so that the force-displacement relation agrees with Hertzian contact theory. The collision time was stretched by using a numerical procedure to solve the resulting equations of the non-linear spring-dashpot oscillator. For the tangential component, they developed a model based on the assumption that, throughout solid contact, a particle either rolls or slides, depending on the particles' incidence angle. Although the approach of considering pure rolling for small incidence angles does not reproduce collisions with recoil of the contact point, their methodology can be easily adapted to account for it. Even though their model is able to reproduce normal and oblique collisions in viscous liquids with satisfactory realism, the fact that it needs an extra iterative procedure to deal with the non-linear spring when computing the normal force may deteriorate its computational performance for denser concentrations. Furthermore, the force law for the tangential component of the collision force depends on the particles' incidence angle, which is difficult to interpret, e.g., for cases in which geometrical constrains force sustained contact.

In the present study we present a new model for wet particle-particle and particle-wall collisions in fully-resolved simulations of particle-laden flows. We show that a simple variant of a linear spring-dashpot model capable of stretching the collision time \cite{VanderHoef2004}, \cite{Breugem2010} suffices for computing contact forces. The contact model can be seen as a linearized version of Hertz contact theory, and its choice is motivated by a separation between the time scales of solid-solid contact and particle motion. The advantage of using this model is that its parameters can be analytically determined from well-documented material properties and a desired collision time, which is computationally attractive. Moreover, it accounts for stick-slip effects at the contact point without requiring explicit definition of impact and rebound angles. Oblique collisions with recoil are explicitly accounted for by using a tangential coefficient of restitution $e_t$ as input parameter. This contact force model is implemented in an efficient and second-order accurate IBM for particle laden flows developed in \cite{Breugem2012} and combined with a physical model for lubrication interactions and roughness effects. We found the experimental data used by \citet{Kempe2012} to validate their approach to be a good set of canonical tests for which a physically realistic collision model should pass. We therefore validated our collision model against those distinct experimental cases which include the trajectory of a sphere colliding onto a planar surface in a viscous liquid \cite{Gondret2002}, head-on particle-particle collisions \cite{Yang2006} and data oblique particle-wall collisions \cite{Joseph2004}. 

This article is organized as follows. Section \ref{sec:cphys} presents a brief overview of the physics of dry collisions of elastic spheres (\ref{subsec:phys}) followed by the description of the methodology for computing contact force/torques (\ref{sec:sscm}). Section \ref{sec:lub} addresses the effects of the interstitial fluid in a wet collision and our modeling strategy for lubrication interactions. The numerical implementation is addressed in Section \ref{sec:nummeth}. Section \ref{sec:results} explores the consequences of excessive and insufficient stretching of the collision time and presents the validation of the model against several benchmark experiments. Finally, in Section \ref{sec:conclu} the conclusions and outlook are given.

\section{Dry Collisions}
\label{sec:cphys}
\subsection{Physics}
\label{subsec:phys}
\begin{figure}[htp!]
\centering
\includegraphics[height=0.35\textwidth]{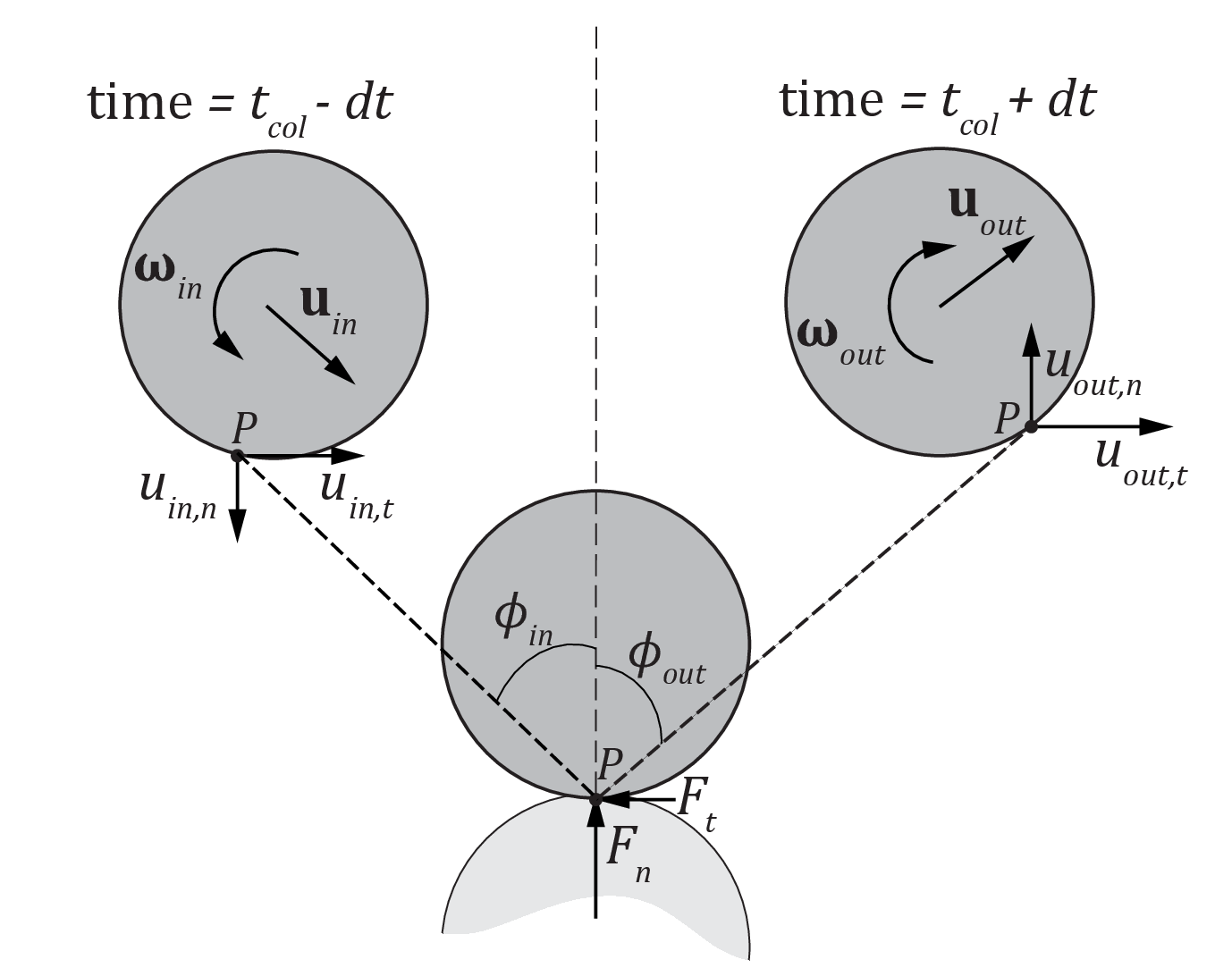}
\caption{Schematic representation of an oblique inter-particle collision. For the sake of clarity we considered in this figure a reference frame moving with the light grey particle, which implies that the velocities sketched are relative velocities. $F_n$ and $F_t$ denote the normal and tangential component of the collision force.}
\label{fig:skpp}
\end{figure}

\begin{figure}[htp!]
\centering
\includegraphics[width=0.45\textwidth]{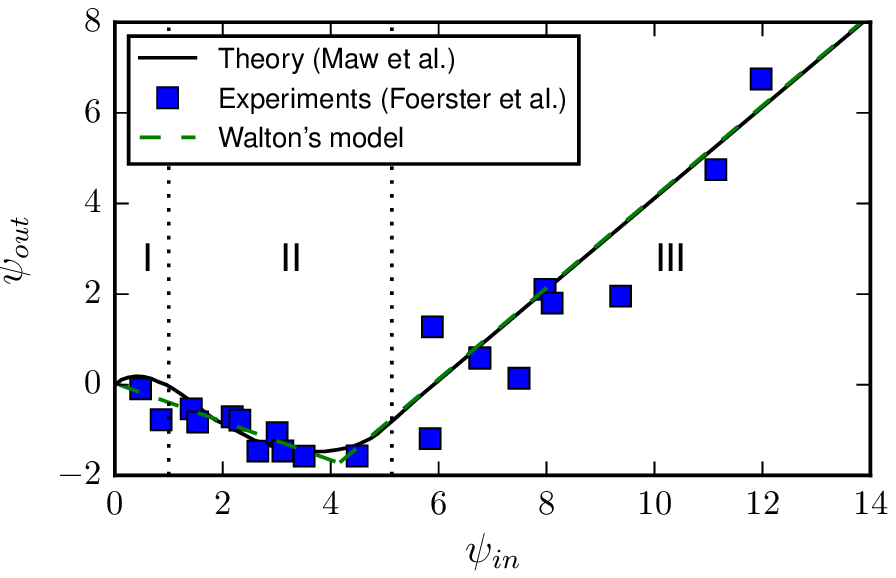}
\caption{(Color online) Numerical solution of the model of \citet{Maw1976} for collisions between glass spheres, compared with experimental data of \citet{Foerster1994} and the model of \citet{Walton1993}. The curve and experimental data were extracted from a curve $\Psi_{out}$ vs $\Psi_{in}$ of \cite{Foerster1994} and rescaled to $\psi_{out}$ vs $\psi_{in}$ with the parameters of their homogeneous $3\mathrm{mm}$ glass spheres, $\nu = 0.22$ and $\mu_c = 0.092$. The two vertical dotted lines delimit the three different types of impact and are given by $\psi_{in} = 1$ and $\psi_{in} = 4\chi-1 \approx 4.2$.}
\label{fig:maws_solution}
\end{figure}

When head-on inter-particle collisions take place in the absence of a viscous fluid, kinetic energy is dissipated exclusively due the contact mechanics. This energy loss can be described by a \textit{dry} coefficient of restitution, $e_{n,d}$, defined as the ratio of the relative rebound velocity to the relative impact velocity. The collision is referred to as oblique when the particles approach each other with an incidence angle just prior to contact $\phi_{in}$ and bounce with a rebound angle $\phi_{out}$, as illustrated in Figure \ref{fig:skpp}.
From these, it is convenient to define \textit{effective} angles of incidence and rebound, respectively as,
\begin{align}
\Psi_{in} &= \frac{u_{in,t}}{u_{in,n}} = \tan(\phi_{in})\mathrm{, \; and} \\
\Psi_{out} &= \frac{u_{out,t}}{u_{in,n}} = e_{n,d}\tan(\phi_{out})\mathrm{,}
\end{align}
with the normal dry coefficient of restitution $e_{n,d}$ given by
\begin{equation}
e_{n,d} = \frac{u_{out,n}}{u_{in,n}}\mathrm{.}
\end{equation}
\citet{Maw1976} explored this problem in detail. They used Hertzian contact theory to obtain the normal component of the collision force and velocity. Moreover, they assumed particles of the same material for which the contact area consists of stick and slip regions, and that slip could be modeled by a constant coefficient of sliding friction, $\mu_c$. Their results show that three different types of impacts can occur, depending on the value of the following normalized incidence angle,
\begin{equation}
\psi_{in} = \frac{2}{\mu_c}\frac{1-\nu}{2-\nu} \Psi_{in}\mathrm{,}
\end{equation}
and the material- and geometry-dependent parameter,
\begin{equation}
\chi = \left(1+\frac{1}{K^2}\right)\frac{1-\nu}{2-\nu}\mathrm{,}
\end{equation}
where $\nu$ is the Poisson's ratio and $K$ the normalized particle radius of gyration ($K^2=2/5$ for a homogeneous solid sphere).
Figure \ref{fig:maws_solution} shows $\psi_{out}$ as a function of $\psi_{in}$ as computed in their model, where $\psi_{out}$ is the normalized rebound angle, defined analogously to $\psi_{in}$ as,
\begin{equation}
\psi_{out} = \frac{2}{\mu_c}\frac{1-\nu}{2-\nu} \Psi_{out}\mathrm{.}
\end{equation}

The numerical solution of this model yields three distinct regions denoted in Figure \ref{fig:maws_solution} by I, II and III. First, for small incidence angles, $\psi_{in} \leq 1$, the sphere sticks during contact because the normal component of the load is much larger than the tangential component. When the contact surface starts to shrink, small regions of micro-slip may occur due to tangential elastic recovery, which can spread throughout the entire contact area, leading to gross slip. Second, for an intermediate range of incidence angles, $1<\psi_{in}\leq4\chi-1$, the collision starts with gross slip, but the frictional stresses retard the tangential velocity, which rapidly drops to zero in the entire contact area (full stick). Finally, for higher incidence angles, $\psi_{in}>4\chi-1$, the tangential component of the load is even higher and gross slip occurs throughout the contact time.

\citet{Walton1993} proposed a simplified hard-sphere model with three parameters: (1) a normal coefficient of restitution, $e_{n,d}$; (2) a tangential coefficient of restitution for non-sliding contact, $e_{t,d}$, defined as,
\begin{equation}
e_{t,d} \equiv - \frac{u_{out,t}}{u_{in,t}} = -\frac{\Psi_{out}}{\Psi_{in}}\mathrm{;}
\end{equation}
and (3) a coefficient of sliding friction, $\mu_c$, to model the tangential component of the load, $F_t$, acting on the particle when it is sliding,
\begin{equation}
F_t=-\mu_c |F_n|\mathrm{,}
\end{equation}
where $F_n$ is the normal component of the contact force acting on the particle. This model assumes that the collision force acts at a single point and can be decomposed into a normal and tangential component. It further assumes that throughout the collision time the regime is either full stick or gross slip.
From these three parameters, one can define the two lines which dictate the collision regime: 
\begin{subequations}
 \begin{empheq}[left={\Psi_{out}=\empheqlbrace}]{align}
    -e_{t,d}\Psi_{in}  &, \Psi_{in} \le \Psi_{in}^* \,\,\, \mathrm{(stick),} \label{eq:walton1st} \\
    \Psi_{in} -\mu_c(1+1/K^2)(1+e_{n,d}) &, \Psi_{in} > \Psi_{in}^* \,\,\, \mathrm{(slip),}\label{eq:walton2nd}
  \end{empheq}
\end{subequations}
where Eq. \eqref{eq:walton1st} is obtained directly from the definition of $e_{t,d}$, and Eq. \eqref{eq:walton2nd} by applying the definition of coefficient of sliding friction to relate the normal and tangential momentum impulses. $\Psi_{in}^*$ is the incidence angle above which the collision regime changes from full stick into gross slip:
\begin{equation}
\Psi_{out}(\Psi_{in}^{*+}) = \Psi_{out}(\Psi_{in}^{*-}) \Leftrightarrow \Psi_{in}^* = \mu_c\left(1+\frac{1}{K^2}\right)\frac{1+e_{n,d}}{1+e_{t,d}}\mathrm{.}\label{eq:psi_crit}
\end{equation}

The two models differ most significantly in the intermediate region of incidence angles, for which there may be periods of full stick and gross slip throughout the contact. Despite these differences, the simplified approach is able to reproduce experimental data reasonably well, as shown, e.g., in \cite{Walton1993}, \cite{Foerster1994} (Figure \ref{fig:maws_solution}) and \cite{Joseph2004}. The minimalistic nature Walton's model makes it an attractive for problems where a detailed description of the contact mechanics is not required, which is in general the case for particle-laden flows.

\subsection{Modeling}
\label{sec:sscm}

\citet{Legendre2006} demonstrated that collisions of spherical particles in a viscous liquid have a contact time larger but of the same order than the contact time in a dry system, predicted by Hertzian contact theory. They show that this contact time is four to five orders of magnitude smaller than the viscous relaxation time of the particle, depending on the impact Stokes number. This means that that the particle experiences a collision as a \textit{discontinuity} in its motion. Even if the characteristic time scale of the particle motion is not dictated by the viscous relaxation time, (e.g., due to geometrical constrains in a flow with high volume fraction of particles) this clear separation of time scales typically remains valid. Hence, we require that the collision dynamics are realistically reproduced from a \textit{macroscale} perspective, i.e., realistic approach and rebound velocities and timescale small enough such that this separation of time scales is satisfied. Hence, it is convenient to use a model capable of stretching the collision time, so that that the overall numerical algorithm is not significantly penalized by the overhead introduced by the integration of the particles' equations of motion. Furthermore, it is convenient to use a model with parameters that can be easily measured experimentally, such as the parameters of Walton's model. \citet{Joseph2004} successfully used this model to describe experimental data from wet oblique collisions of spherical particles onto planar surfaces, which further supports its validity to describe collisions in a viscous liquid.

\begin{figure}[htp!]
   \subfigure[]{
        \centering
        \includegraphics[width=0.35\columnwidth]{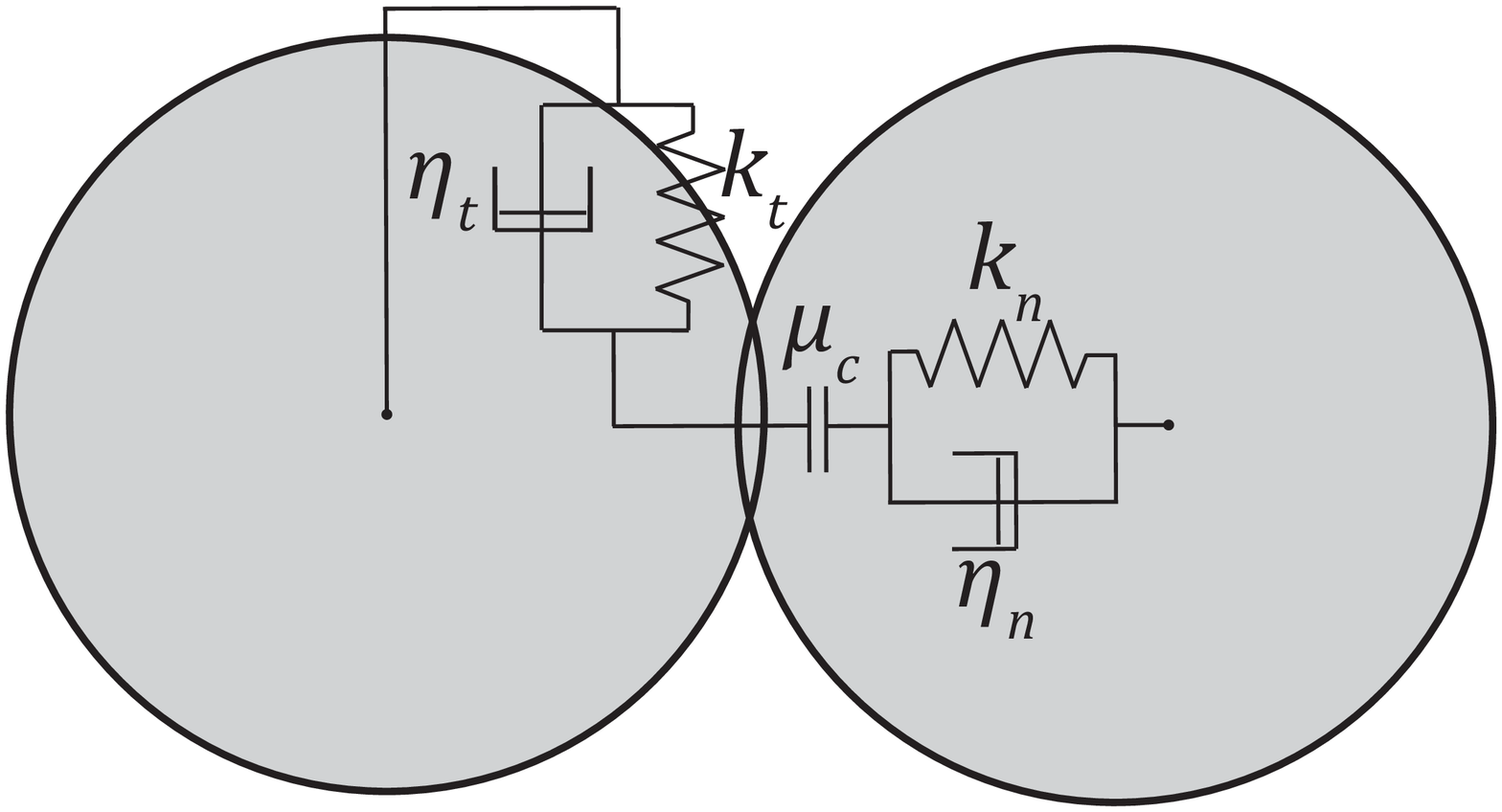}
        \label{fig:sscm}}
\quad\quad
   \subfigure[]{
        \centering
        \includegraphics[width=0.35\columnwidth]{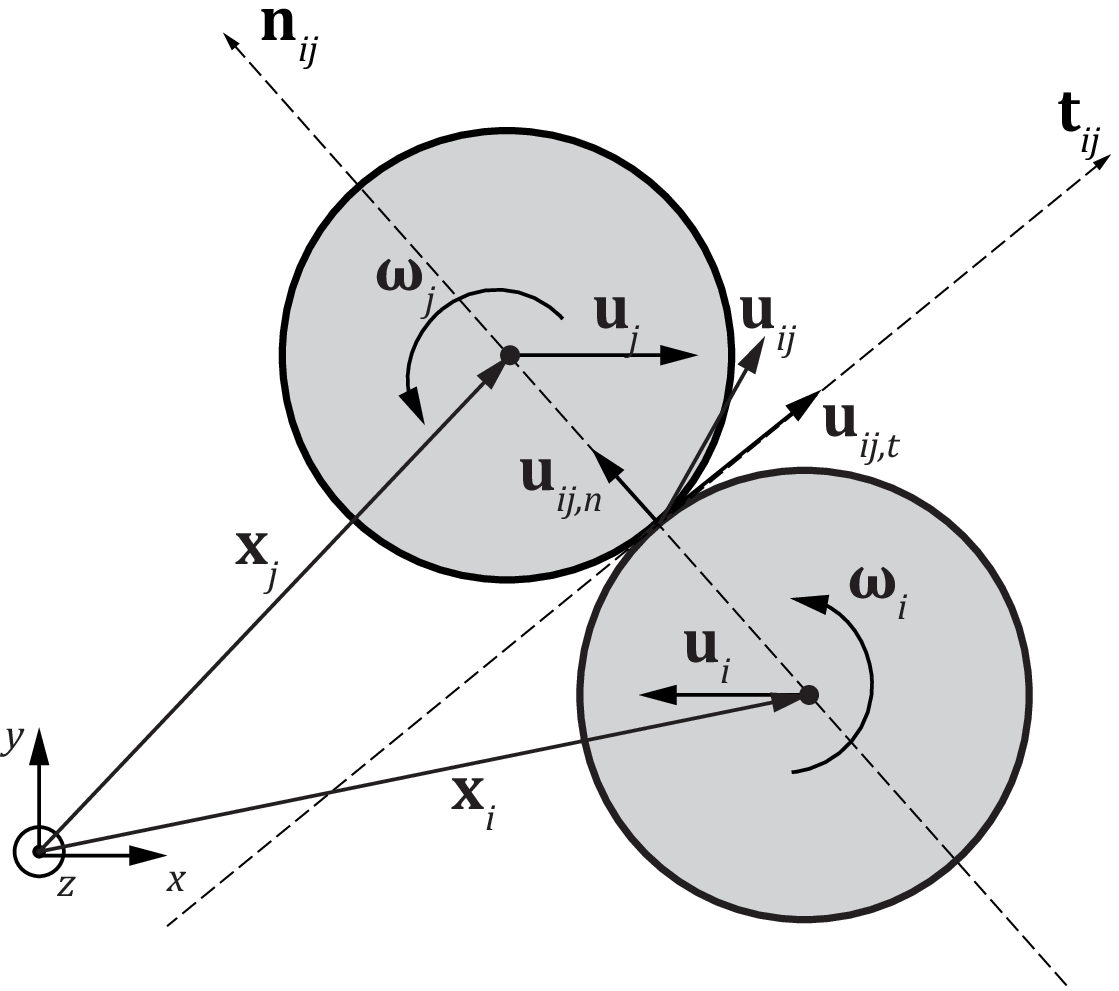}
        \label{fig:oblique_col}}
   \caption{(a) Linear spring-dashpot model. (b) Notation and reference frame adopted for an inter-particle collision.}
   \label{fig:sscm_explained}
\end{figure}

We found the variant of the soft-sphere contact model of \citet{Tsuji1993}, described in \cite{VanderHoef2004} to be suitable for our problem. This approach has computational advantages for dense suspensions when compared to other alternatives such as hard-sphere models and allows the collision time to be stretched. The model consists on a linear spring-dashpot system in the normal and tangential directions, with a Coulomb friction slider in the latter, as sketched in Figure \ref{fig:sscm}. In the following we describe the model, with differences in the definition of the tangential unit vector and in the value to which the tangential displacement is saturated. Figure \ref{fig:oblique_col} illustrates the notation and reference frame adopted.

The normal force acting on particle $i$ due to a contact with particle $j$, with a relative velocity at the contact point given by
\begin{equation}
\mathbf{u}_{ij} = \left(\mathbf{u}_i+R_i\boldsymbol{\omega}_i\times \mathbf{n}_{ij}\right) -\left(\mathbf{u}_j+R_j\boldsymbol{\omega}_j\times \mathbf{n}_{ji}\right)\mathrm{,}
\label{eq:deltau}
\end{equation}
is the component of the collision force that acts along the direction of the line-of-centers (Figure \ref{fig:oblique_col}),
\begin{equation}
\mathbf{n}_{ij} = \frac{\mathbf{x}_j-\mathbf{x}_i}{||\mathbf{x}_j-\mathbf{x}_i||}\mathrm{.}
\end{equation}
This collision force depends on the overlap distance between the two particles,
\begin{equation}
\boldsymbol{\delta}_{ij,n} = \left(R_i+R_j-||\mathbf{x}_j-\mathbf{x}_i||\right)\mathbf{n}_{ij}\mathrm{,}\label{eq:deltan}
\end{equation}
and the normal relative velocity of the contact point, 
\begin{equation}
\mathbf{u}_{ij,n}=(\mathbf{u}_{ij}\cdot \mathbf{n}_{ij})\mathbf{n}_{ij} \mathrm{,}
\end{equation}
and is obtained from the equivalent linear spring-dashpot system:
\begin{equation}
\mathbf{F}_{ij,n} = -k_n\boldsymbol{\delta}_{ij,n}-\eta_n\mathbf{u}_{ij,n}\mathrm{,}
\label{eq:normf}
\end{equation}
where $k_n$ and $\eta_n$ are the normal spring and dashpot coefficients, respectively. These are computed by solving for the motion of a linear harmonic oscillator \cite{VanderHoef2004}, and requiring that there is no overlap at the end of the collision, $t=N\Delta t$,
\begin{equation}
\left(\boldsymbol{\delta}_{ij,n}\cdot \mathbf{n}_{ij}\right) |_{t=N\Delta t} = 0\mathrm{,} 
\label{eq:ic1}
\end{equation}
and that the velocity at the end of the collision is given by the definition of $e_{n,d}$,
\begin{equation}
\left(\mathbf{u}_{ij,n}\cdot\mathbf{n}_{ij} \right)|_{t=N\Delta t} = -e_{n,d} \left(\mathbf{u}_{ij,n}\cdot\mathbf{n}_{ij}\right)|_{t=0}\mathrm{.}
\label{eq:ic2}
\end{equation}
Note that we define the collision time, $T_n$, as a multiple $N$ of the time step of the overall numerical algorithm, $\Delta t$. This is convenient because -- as our results will show --  the outcome of a numerical simulation of a wet collision is more realistic if the fluid is allowed to adapt itself to the sudden changes in particle velocity. In practice, because $T_n$ should be fixed during a collision, and $\Delta t$ may vary in agreement with the stability criterion of the fluid solver, one should define the collision time as a multiple of the \textit{estimated} time step of the numerical algorithm.

The coefficients read,
\begin{equation}
k_n = \frac{m_e\left(\pi^2+\ln^2 e_{n,d}\right)}{(N\Delta t)^2}\mathrm{,} \; \eta_n = -\frac{2m_e\ln e_{n,d}}{(N\Delta t)}\mathrm{,} 
\end{equation}
where
\begin{equation}
m_{e} = \left(m_i^{-1}+m_j^{-1}\right)^{-1}\mathrm{,}
\end{equation}
is the reduced mass of the particles.

This approach can be seen as a linearized version of Hertzian contact theory. Since we model the collision as a discontinuity in the particle motion, it is sufficient to guarantee that the conditions specified in equations \eqref{eq:ic1} and \eqref{eq:ic2} are fulfilled and $N\Delta t$ is small enough so that the separation of time scales is satisfied in good approximation. One advantage of using a linear system is that the spring and dash-pot constants can be determined analytically and a priori, which is computationally attractive. Notice that increasing the value of $T_n$ reduces the spring stiffness, which makes the contact \textit{softer}. This implies that excessive stretching of the collision time results in a large overlap between solid surfaces and consequently in a unrealistic delay of the particle rebound. On the other hand, the collision time should be sufficiently stretched so that the collision force is accurately resolved in time. We require that the maximum particle overlap, which is reached when the particles have zero relative velocity, $\delta_{ij,n}^{max}=\delta_{ij,n}|_{u_{ij,n}=0}$, is much smaller than $D_p$:
\begin{equation}
T_n \ll T_{n}^{*} = a\frac{D_p}{{(\mathbf{u}_{ij}}\cdot \mathbf{n}_{ij})|_{t=0}}\mathrm{e}^{-\left(\arcsin(\pi/a)/\pi\right)} \label{eq:deltamax_best}
\end{equation}
where $a = \sqrt{\pi^2+\ln^2(e_{n,d})}$ \cite{VanderHoef2004}. Alternatively, if applicable, one can require that the maximum overlap due to the particle's submerged weight ($\delta_{ij,n}^{max,g} = |1-\rho_f/\rho_p|g/k_n$) is much smaller than $D_p$:
\begin{equation}
T_{n} \ll T_{n}^{*,g} =  \sqrt{\frac{D_p}{g}\frac{a^2}{|1-\rho_f/\rho_p|}}\mathrm{.} \label{eq:deltamax}
\end{equation}

The tangential force is obtained analogously to $\mathbf{F}_{ij,n}$, but now with a Coulomb friction model to account for sliding motion:
\begin{equation}
\mathbf{F}_{ij,t} = \min\left(||-k_t\boldsymbol{\delta}_{ij,t}-\eta_t\mathbf{u}_{ij,t}||,||-\mu_c \mathbf{F}_{ij,n}||\right)\mathbf{t}_{ij} \label{eq:tangf}
\end{equation}
\noindent where $\boldsymbol{\delta}_{ij,t}$ is the tangential displacement and $\mathbf{t}_{ij}$ the unit vector with the direction of the test force: 
\begin{equation}
\mathbf{t}_{ij} = -\frac{k_t\boldsymbol{\delta}_{ij,t}+\eta_t\mathbf{u}_{ij,t}}{||k_t\boldsymbol{\delta}_{ij,t}+\eta_t\mathbf{u}_{ij,t}||}\mathrm{.}
\end{equation}
The coefficients $k_t$ and $\eta_t$ are obtained in an analogous way by solving an harmonic oscillator for the tangential direction, and requiring that the definition of the tangential coefficient of restitution is fulfilled,
\begin{equation}
\left(\mathbf{u}_{ij,t}\cdot\mathbf{t}_{ij}\right) |_{t=N\Delta t} = -e_{t,d}\left(\mathbf{u}_{ij,t}\cdot\mathbf{t}_{ij}\right)|_{t=0}\mathrm{,}
\end{equation}
and that the collision times in the normal and tangential directions match ($T_t = T_n$).
The values of the coefficients read,
\begin{equation}
k_t = \frac{m_{e,t}\left(\pi^2+\ln^2 e_{t,d}\right)}{(N\Delta t)^2} \mathrm{,} \;\; \eta_t = -\frac{2 m_{e,t}\ln e_{t,d}}{(N\Delta t)}\mathrm{,}
\label{eq:tangcoeffs}
\end{equation}
where the reduced mass of the system is given by:
\begin{equation}
m_{e,t} = \left(1+1/K^2\right)^{-1}m_e\mathrm{.}
\label{eq:met}
\end{equation}
The tangential displacement of the contact point must be integrated in time from the imminence of contact. From the integration of the relative tangential velocity at the point of contact we get
\begin{equation}
\boldsymbol{\delta}_{ij,t}^{*n+1} = \underline{\mathbf{R}} \cdot \boldsymbol{\delta}_{ij,t}^{n} + \int_{t^n}^{t^{n+1}} \! \mathbf{u}_{ij,t} \, \mathrm{d}t\mathrm{,}
\label{eq:deltat1}
\end{equation}
where $\underline{\mathbf{R}}$ is a rotation tensor which rotates $\boldsymbol{\delta}_{ij,t}^{n}$ to the new local coordinate system at time level $n+1$.

The tangential force becomes independent of the tangential displacement of the spring when the particle starts sliding (Eq. \eqref{eq:tangf}). If the tangential displacement is further incremented when the particle starts to slide, unrealistic results can be obtained if the collision regime changes subsequently to sticking \cite{Brendel1998}. Hence, the tangential displacement must be saturated in order to comply with Coulomb's condition, whenever the collision is in the sliding regime \cite{Luding2008}:
\begin{subequations}
 \begin{empheq}[left={\boldsymbol{\delta}_{ij,t}^{n+1}=\empheqlbrace}]{align}
    \boldsymbol{\delta}_{ij,t}^{*n+1} \,\, &, ||\mathbf{F}_{ij,t}|| \le \mu_c ||\mathbf{F}_{ij,n}|| \mathrm{,} \\ 
    (1/k_t)\left(-\mu_c||\mathbf{F}_{ij,n}||\mathbf{t}_{ij}-\eta_t\mathbf{u}_{ij,t}\right) \,\, &, ||\mathbf{F}_{ij,t}|| > \mu_c ||\mathbf{F}_{ij,n}|| \mathrm{.}
 \end{empheq}
 \label{eq:deltat}
\end{subequations}

After computing the contact forces acting at the point of contact, we determine the equivalent force and couple acting in the particle centroid:
\begin{align}
\mathbf{F}_{ij}^c &= \mathbf{F}_{ij,t}+\mathbf{F}_{ij,n}\mathrm{,}\label{eq:ftot} \\
\mathbf{T}_{ij}^c &= R_p\left(\mathbf{n}_{ij}\times\mathbf{F}_{ij,t}\right)\mathrm{.} \label{eq:ttot}
\end{align}
The total collision force and torque are the sum of contributions of all the particles in direct contact with the particle $i$:
\begin{subequations}
 \begin{empheq}[]{align}
  \mathbf{F}_i^c = \sum_j \mathbf{F}_{ij}^c \mathrm{,} \\ 
  \mathbf{T}_i^c = \sum_j \mathbf{T}_{ij}^c\mathrm{.} \label{eq:ftsum}
 \end{empheq}
\end{subequations}

A wall is treated as a semi-infinite spherical particle, which makes a particle-wall collision the limit case of a spherical particle with finite-radius, $R_{i}$, colliding onto a sphere with radius $R_{j}\rightarrow \infty$. Thus, the parameters for particle-wall collisions are computed in a similar way by taking this limit. The reduced mass is now given by $m_e = m_i$ and  the normal overlap by $\boldsymbol{\delta}_{iw,n} = \left(R_i-||\mathbf{x}_i-\mathbf{x}_{w}||\right)\mathbf{n}_{iw}$ where $\mathbf{n}_{iw}$ is the unit-vector perpendicular to the wall and $\mathbf{x}_{w}$ the coordinate of the point of contact on the planar surface.

\section{Effects of the interstitial fluid}
\label{sec:lub}

\subsection{Lubrication effects}
\label{subsec:lub_eff}

A particle immersed in a viscous liquid experiences lubrication effects when moving close to and with finite relative velocity to another particle or wall. Assuming a drainage of the intervening liquid film in the Stokes regime, the force acting on the particle has an analytical solution that diverges when the non-dimensional gap-width, $\varepsilon \equiv \delta_{ij,n}/R_p$, tends to zero \cite{Brenner1961}. Our IBM is able to reproduce this and other analytical solutions until a certain (small) value of $\varepsilon$. For smaller gap-widths ($\lesssim \Delta x$) the IBM under-predicts this force due to a lack of spatial grid resolution. An approach that has been adopted for these cases is to keep the grid fixed and use lubrication models based on asymptotic expansions of analytical solutions for the lubrication force in the Stokes regime to compensate this lack of spatial grid resolution (e.g. \cite{Ten2002}, \cite{Breugem2010}, \cite{Kempe2012} and \cite{Motta2013}). Taking these effects into account has been proved to be important for computing realistic bouncing velocities in simulations of head-on particle-wall collisions in viscous fluids \cite{Kempe2012}.

Lubrication theory shows that ideally smooth particles would not reach actual solid-solid contact. Even if one accounts for the particles' surface deformation due to the abrupt increase of the pressure in the gap, the particles would not reach direct contact but a finite \textit{closest distance of approach}, $h_m$ \cite{Davis1986}. However, particles may interact through their asperities, with typical size $\sigma$, before reaching a gap distance of $h_m$. \citet{Joseph2001} observed larger scatter of their experimental data for wet head-on collisions of a spherical particle onto a planar surface  when $\sigma>h_m$. They argued that the contact occurs through the asperities, which are irregularly oriented, before elastohydrodynamic lubrication effects become important. This reasoning validates the approach used by several authors of setting the lubrication correction to zero for small gap-widths (e.g., \cite{Ten2002}, \cite{Kempe2012}) or making it independent of the gap-width \cite{Breugem2010}.

The most important component of the lubrication forces acting on the particle is the squeezing force acting along the line-of-centers, because its dominant term is $\propto 1/\varepsilon$ in contrast to translational and rotational shearing, which diverge slower ($\propto \ln \varepsilon$) and even for a value of $\varepsilon$ compliant with surface roughness have a negligible effect in the particle dynamics. Test simulations showed that the latter mentioned lubrication corrections had little effect on the results for immersed oblique, particle-wall collisions and therefore we decided to neglect them in the present study.

We use a two-parameter model to account for normal lubrication interactions and roughness effects, as illustrated in Figure \ref{fig:lub_model}. When a spherical particle approaches a planar surface/another particle, for a certain gap-width, $\varepsilon_{\Delta x}$, the IBM cannot resolve the lubrication force acting on the particle. Hence, for gap-widths smaller than $\varepsilon_{\Delta x}$, we correct the lubrication force acting on the particle by adding to the Newton-Euler equations $\Delta F_{lub} = -6\pi\mu R_p u_{ij,n}(\lambda({\varepsilon})-\lambda(\varepsilon_{\Delta x}))$, where the Stokes amplification factor $\lambda$ is given by \cite{Jeffrey1982}:
\begin{figure}[htp!]
\centering
\includegraphics[width=0.7\textwidth]{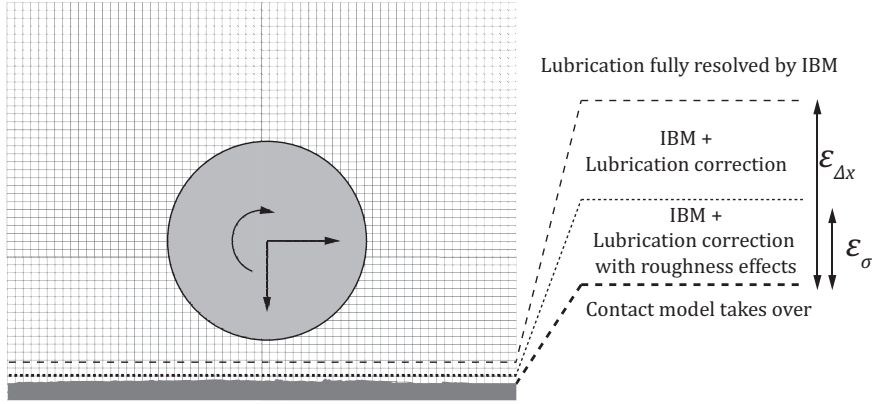}
\caption{Schematic representation of the lubrication model. We illustrate the case of particle-wall interactions for the sake of simplicity. The model is analogous for particle-particle interactions.}
\label{fig:lub_model}
\end{figure}
\begin{alignat}{5}
\lambda_{pp}(\varepsilon) &= \frac{1}{2\varepsilon} &{}- \frac{9}{20}\ln\varepsilon &{}- \frac{3}{56}\varepsilon\ln\varepsilon &{}+ {\cal O}(1)\mathrm{,} \\
\lambda_{pw}(\varepsilon) &= \frac{1}{\varepsilon} &{}- \frac{1}{5}\ln\varepsilon &{}- \frac{1}{21}\varepsilon\ln\varepsilon &{}+ {\cal O}(1)\mathrm{,}
\end{alignat}
for lubrication interactions between two equal spheres, and between a sphere and a planar surface, respectively.

\begin{figure}[htp!]
   \subfigure[]{
   \centering
   \includegraphics[width=0.45\columnwidth]{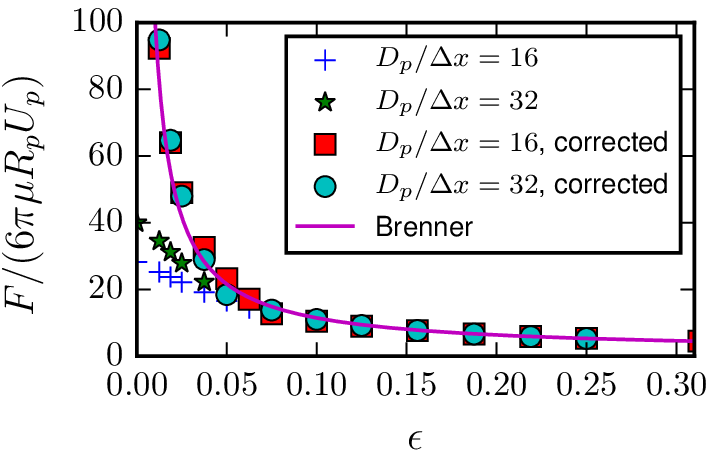}}
   \quad
   \subfigure[]{
   \centering
   \includegraphics[width=0.45\columnwidth]{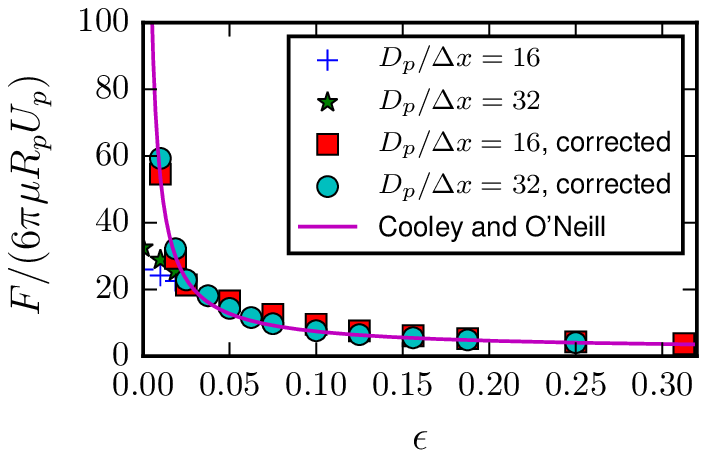}}
   \caption{(Color online) Lubrication corrections for the cases of normal particle-wall (a) and particle-particle (b) interactions, respectively compared against the analytical solution of \citet{Brenner1961} and \citet{Cooley1969}. Results shown for two different resolutions, $D_p/\Delta x = 16$ and $32$.}
   \label{fig:lub_wp}
\end{figure}

The value of $\varepsilon_{\Delta x}$ can be determined by simulating the slow approach of a sphere towards a planar surface \cite{Brenner1961} or between two spheres \cite{Cooley1969} and determining up to which point the IBM is able to reproduce the lubrication interaction \cite{Breugem2010}.
We illustrate this in Figure \ref{fig:lub_wp} by comparing the analytical solution with the simulations without lubrication correction and with lubrication correction. The corresponding values of $\varepsilon_{\Delta x}$ for two different spatial resolutions are given in Table \ref{tbl:lubparam_st}.
\begin{table}[bbh]
\caption{Parameters for the lubrication model}
\centering
\begin{tabular}{clc}
\hline
\hline
$D_p/\Delta x$ & Interaction & $\varepsilon_{\Delta x}$ \\
\hline
$16$ & particle-wall & $0.075$ \\
$16$ & particle-particle & $0.025$ \\
$32$ & particle-wall & $0.05$ \\
$32$ & particle-particle & $0.025$ \\
\hline
\hline
\end{tabular}
\label{tbl:lubparam_st}
\end{table}
To account for surface roughness, we saturate the Stokes amplification factor for gap-widths below a threshold $\varepsilon_\sigma$ so that $\lambda(\varepsilon<\varepsilon_\sigma) = \lambda(\varepsilon_\sigma)$. This threshold value is related to the typical size of the asperities and was fixed to $\varepsilon_\sigma = 0.001$ for particle-wall interactions. We keep the Stokes amplification factor saturated until the surfaces overlap; then the collision force takes over. Hence, the force acting on the particle is corrected by $\Delta F_{lub}$, given by:

\begin{align}
\frac{\Delta F_{lub}}{-6\pi\mu R_p u_{ij,n}} =
\begin{cases}
\lambda(\varepsilon) \,\,\,- \lambda(\varepsilon_{\Delta x}) , \,\,\, &\varepsilon_{\sigma} \leq \varepsilon < \varepsilon_{\Delta x} \\
\lambda(\varepsilon_\sigma) - \lambda(\varepsilon_{\Delta x}) , \,\,\, & 0 \leq \varepsilon < \varepsilon_{\sigma} \\
0, \,\,\, &\mathrm{otherwise.}
\end{cases}
\end{align}

For particle-wall collisions, the normal fluid-induced forces are set to zero for overlaps larger than the overlap due to the particle's submerged weight, $\delta_{ij,n}^g = |\rho_p-\rho_f|g V_p/k_n$, in order to avoid artificial dissipation due to the stretching of the collision time of the contact model. This procedure is not extended to particle-particle interactions, as it can cause significant artificial increase in the particles' acceleration for colliding particle pairs due to a sudden decrease in drag force.

\subsection{Piezoviscous effects}

\citet{Joseph2004} performed experiments on wet, oblique collisions of spheres onto planar surfaces. They showed that the coefficient of sliding friction decreased by one order of magnitude when their smooth steel spheres collide, whereas it remained of the same order of magnitude ($\sim 15\%$ higher) for the case of rough glass spheres. They suggested that this abrupt decrease of the friction coefficient for smooth spheres was due to the fact that a characteristic piezoviscous lenghtscale \cite{Barnocky1988}, $h_{pv}$, was larger than the average size of the asperities and therefore 'contact' occurs through the fluid, which is behaving like an elastic-solid. Also, the slight increase in the coefficient of sliding friction for rough spheres was explained by the fact that the fluid introduces an extra resistance when the asperities have relative motion in the tangential direction. They developed a model capable of predicting the coefficient of sliding friction of smooth spheres colliding onto planar surfaces in a viscous liquid from elastohydrodynamic lubrication theory.

Hence, for the case in which piezoviscous effects are important, it does not suffice to use input parameters from dry collisions and lubrication corrections for obtaining a physically realistic result: the coefficient of sliding friction measured in a wet collision experiment, or predicted by the model developed in \cite{Joseph2004}, $\mu_{c,wet}$, should be used.

\section{Numerical implementation}
\label{sec:nummeth}

The fluid phase is governed by the Navier-Stokes equations for an incompressible Newtonian fluid:
\begin{subequations}
\begin{align}
\boldsymbol{\nabla} \cdot {\bf u} & = 0 \mathrm{,} \label{eq:gov_cont} \\
\frac{\partial {\bf u}}{\partial t} + \boldsymbol{\nabla} \cdot {\bf u}{\bf u} & = - \frac{1}{\rho_f}{\boldsymbol{\nabla} p} + \nu_f {\boldsymbol{\nabla}^2 {\bf u}}\mathrm{,} \label{eq:gov_mom}
\end{align}
\end{subequations}
where ${\bf u}$ is the fluid velocity, $\nu_f \equiv \mu/\rho_f$ the kinematic viscosity of the fluid and $p$ the pressure.

The translational and rotational motion of solid particles is described by the Newton-Euler equations for rigid body motion. For a spherical particle they read,
\begin{subequations}
\begin{align}
\rho_p V_p \frac{\diff \mathbf{u}_c}{\diff t} &= \oint_{\partial V} \! \boldsymbol{\tau} \cdot \mathbf{n} \, \diff A + (\rho_p-\rho_f)V_p\mathbf{g}+ \mathbf{F}_c \mathrm{,} \label{eq:gov_newton} \\
I_p \frac{\diff \boldsymbol{\omega}_c}{\diff t} &=\oint_{\partial V} \! \mathbf{r} \times (\boldsymbol{\tau} \cdot \mathbf{n}) \, \diff A + \mathbf{T}_c \mathrm{.} \label{eq:gov_euler}
\end{align}
\end{subequations}
The left-hand-side of \eqref{eq:gov_newton} is the temporal variation of linear momentum of the particle: $\bf{u}_c$ is the centroid velocity, $\rho_p$ the particle mass density, and $V_p$ the particle volume given by $(4/3)\pi R_p^3$ for a spherical particle with radius $R_p$.

The first term of the right-hand-side of \eqref{eq:gov_newton} is the net force resulting from the distribution of fluid stress, $\mathbf{\tau} \equiv -p \bf{I} + \mu\left(\nabla {\bf u} + \nabla {\bf u}^T \right)$ at the particle surface, $\partial V$, projected to the outward-pointing unit normal to $\partial V$, $\bf{n}$. The second term is the buoyancy force due to a difference between the fluid and particle densities in the presence of a gravitational field with acceleration $\bf{g}$. $\mathbf{F}_c$ represents other external forces acting on the particle such as, e.g., collision forces.

The left-hand-side of \eqref{eq:gov_euler} is the temporal variation of angular momentum of the particle, where $\boldsymbol{\omega}_c$ is its angular velocity and $I_p$ the moment of inertia of a solid sphere, given by $(2/5) \rho_p V_p R_p^2$. Due to spherical symmetry, only two non-trivial terms balance the left-hand-side: the flow-induced torques and the external torques (e.g., a collision torque) respectively the first and second terms in the right-hand-side of \eqref{eq:gov_euler}. $\bf{r}\equiv \bf{x}-\bf{x}_c$ is the position vector relative to the particle centroid $\bf{x} = \bf{x}_c$. $\bf{T}_c$ is an external torque that acts on the particle whenever there is a contact force with a tangential component.

The equations \eqref{eq:gov_cont}, \eqref{eq:gov_mom} and \eqref{eq:gov_newton}, \eqref{eq:gov_euler} form a set of equations coupled through no-slip and no-penetration (ns/np) boundary conditions at the particle/fluid interface. Hence, the velocity at the particle surface,
\begin{equation}
{\bf U}_p  =  {\bf u}_c + {\boldsymbol{ \omega}}_c \times {\bf r}\mathrm{,} \label{eq:vel_surf_part}
\end{equation}
is required to match the local fluid velocity:
\begin{equation}
{\bf u}  =  {\bf U}_p \left( {\bf x} \right)  \;\;\; {\forall} \;\;\; {\bf x} \in \partial V \mathrm{.} \label{eq:ns_np}
\end{equation}

The governing equations for the fluid phase are integrated in time with an explicit low-storage three-step Runge-Kutta method for all terms except the pressure gradient in the Navier-Stokes equations; for the latter the Crank-Nicolson scheme is used. The equations are discretized in space on a uniform, staggered Cartesian grid with the finite-volume method in which spatial derivatives are estimated with the central-differencing scheme. To enforce ns/np conditions at the particle's surface we use the second-order accurate IBM developed in \cite{Breugem2012}. The advantage of an IBM is that the governing equations are solved on a spatially continuous grid without any holes,  which enables the use of an efficient, FFT-based, direct solver for the pressure Poisson equation. Stability restrictions for the computational time step have been derived by \citet{Wess2001}. For a uniform Eulerian grid with $\Delta x = \Delta y = \Delta z$ and the central-differencing scheme, a sufficient criterion for von Neumann stability is given by:
\begin{equation}
\mathrm{Cou} = \frac{\Delta t}{\min \left( \frac{1.65}{12} \frac{\Delta x^2}{\nu_f} , \frac{\sqrt{3} \Delta x}{\sum_{i=1}^3 \left| u_i \right| } \right)} < 1. \label{v2}
\end{equation}

The governing equations for the solid particles are advanced in time with the same Runge-Kutta scheme as used for the fluid phase, except for collision forces/torques and tangential displacement; these terms are integrated with a second-order Crank-Nicolson (CN2) scheme that has proven to return a stable and accurate integration. This scheme requires the contact force at the next time level, $q$, which depends on the values of the particle position and velocity at the same level (Eqs. \eqref{eq:normf} and \eqref{eq:tangf}). We therefore compute the contact force iteratively as a function of the particle position and velocity at $q$ until the new particle position converges. The particles' position and velocity are initialized ($k=0$) with the values of the previous time level $q-1$. The advancement follows directly the integration of the Navier-Stokes equations, within the RK3 time advancement loop with a time step $\Delta t_{p}$ which is allowed to be smaller than the time step of the Navier-Stokes solver $\Delta t$ to ensure that the contact forces and lubrication force corrections are accurately integrated. The forces induced by the IBM are fixed in time while the sub-integrations are performed. For sub-stepping ratios $r_{\Delta t} = \Delta t/\Delta t_{p}$ ranging from $1$ to ${\cal O}(100)$, the extra overhead introduced by the sub-stepping is negligible. The scheme is illustrated below.

{\small
\begin{algorithmic}
\State $k = 0$
\Do
\ForAll {particles $j$ in contact with particle $i$}
\State compute $\boldsymbol{\delta}_{ij,n}^{q,k}$ and $\boldsymbol{\delta}_{ij,t}^{q,k} = \underline{\mathbf{R}} \cdot \boldsymbol{\delta}_{ij,t}^{q-1} + \frac{\Delta t_{p}^q}{2}(\mathbf{u}_{ij,t}^{q,k}+\underline{\mathbf{R}} \cdot \mathbf{u}_{ij,t}^{q-1})$
\State compute $\mathbf{F}_{ij,n}^{q,k}$ and $\mathbf{F}_{ij,t}^{q,k}$
\State update $\mathbf{F}_{c}^{q,k}$ and $\mathbf{T}_{c}^{q,k}$
\EndFor
\begin{flalign}
 & \;\;\:\:\:\,\,{\bf u}^{q,k}_c = {\bf u}_c^{q-1} + \text{(particle-fluid coupling terms \cite{Breugem2012})} + \frac{\Delta t_{p}^q}{2} \frac{{\bf F}_c^{q,k} + {\bf F}_c^{q-1} }{\rho_p V_p} & \label{v3a} \\
 & \;\;\:\:\:\,\,{\bf x}^{q,k}_c = {\bf x}_c^{q-1} + \frac{\Delta t_{p}^q}{2} \left( {\bf u}^{q,k}_c + {\bf u}_c^{q-1} \right) \label{v3b} \\
 & \;\;\:\:\:\,\,\boldsymbol{\omega}^{q,k}_c = {\boldsymbol \omega}_c^{q-1} + \text{(particle-fluid coupling terms \cite{Breugem2012})} + \frac{\Delta t_{p}^q}{2} \frac{{\bf T}_c^{q,k} + {\bf T}_c^{q-1} }{\rho_p I_p} & \label{v1o} \\
 & \;\;\:\:\:\,\,err_{iter}^k = ||{\bf x}^{q,k}_c-{\bf x}^{q,k-1}_c|| 
\end{flalign}
\State $k = k+1$ 
\doWhile{$err_{iter}^k<err_{iter,max}$}
\end{algorithmic}}

$\Delta t_{p}^q = (\alpha_{r} + \beta_{r}) \Delta t_{p}$ varies according to the duration of the Runge-Kutta sub steps and the coefficients can be found in \citet{Wess2001}: $\alpha_1 = 32/60$, $\beta_1 = 0$, $\alpha_2 = 25/60$, $\beta_2 = -17/60$, $\alpha_3 = 45/60$, $\beta_3 = -25/60$. The lubrication corrections are integrated with the same scheme as the collision force. From this CN2 scheme, we expect second-order accuracy for the linear momentum of the particle and consequently, third-order accuracy for the integration of the particle velocity. We verified the accuracy of the method by reproducing, in simulations of dry collisions, the coefficients of restitution ($e_{n,d}$ and $e_{t,d}$) that are used as an input in the collision model (not shown). For the simulations of the present work $1$ iteration sufficed for obtaining a small iterative error: $err_{iter}^1 > 10^{-8}\Delta x$. 

Unless otherwise stated, the particles are resolved with $D_p/\Delta x=16$ and a sub-stepping ratio of $r_{\Delta t}=50$, the collision time set to $T_n = 8\Delta t$ and the time step set by $\mathrm{Cou}=0.5$.

\section{Results from collision simulations}
\label{sec:results}

\subsection{Bouncing motion of a solid sphere colliding onto a planar surface in a viscous liquid}
\label{subsec:headon}
We simulated the bouncing motion of a solid sphere immersed in a viscous liquid and colliding under gravity onto a planar surface. The trajectory of the point of the particle closest to the surface and time evolution of its velocity are compared to the experimental data of \citet{Gondret2002}. This experiment is a useful benchmark for confirming that the lubrication corrections and collision model return a realistic bouncing velocity, and that the collision is represented in good approximation as an instantaneous event in the particle motion. Furthermore, there is no need for specifying impact and rebound velocities, which definitions vary significantly in literature \cite{Izard2014}. Note that small differences in rebound velocity are amplified after its temporal integration, and therefore more noticeable in the particle trajectory. Hence, a good agreement with this experiment gives a fine indication that the approach used to resolve a head-on wet collision is adequate.

The simulations were carried out in a domain corresponding to a closed container with dimensions $L_x/D_p\times L_y/D_p\times L_z/D_p = 12\times 30\times 12$. The particle is initially placed at $y/D_p=L_y-1.5R_p$, centered in $L_x/2$ and $L_z/2$. The motion is driven by a downward-pointing gravitational acceleration of $g = 9.81 \mathrm{m/s^2}$. The time step was fixed to the maximum allowed by the stability criterion at the maximum particle velocity (i.e., at impact), multiplied by $\mathrm{Cou}$ to ensure a stable and accurate temporal integration. The physical and computational parameters are listed in Table \ref{tbl:gondrparam}.

\begin{figure}[htp!]
        \centering
   \subfigure[]{
        \centering
        \includegraphics[width=0.45\columnwidth]{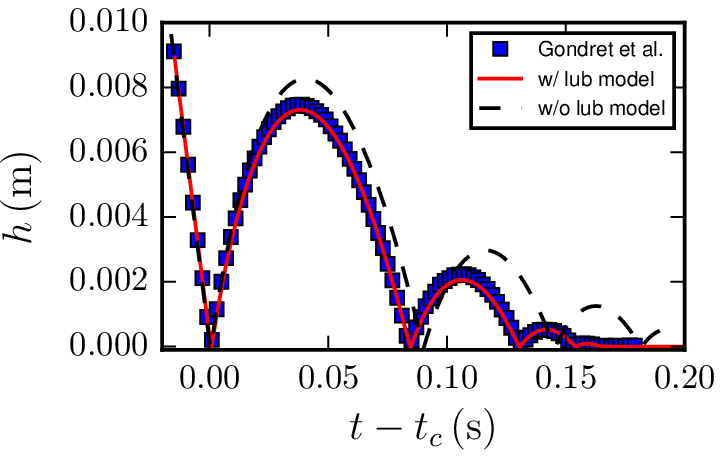}
        \label{fig:bouncing_all}}
\quad
   \subfigure[]{
        \centering
        \includegraphics[width=0.45\columnwidth]{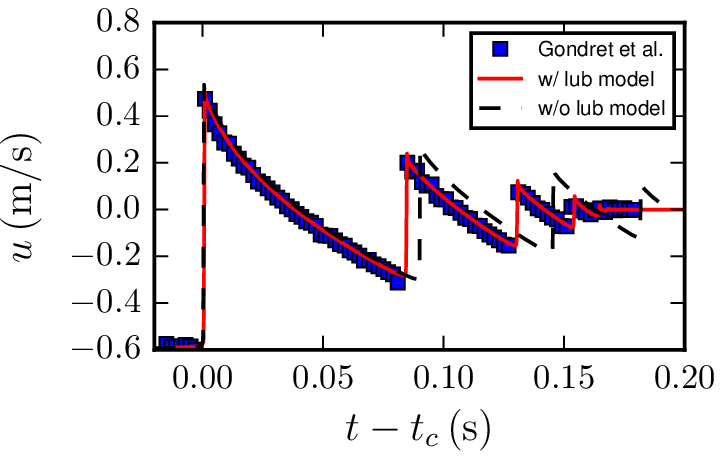}}
   \caption{(Color online) Trajectory (a) and time evolution of the particle velocity (b) in the bouncing motion of a steel sphere colliding onto a planar surface in silicon oil RV10.}
\end{figure}

Figure \ref{fig:bouncing_all} presents the results for the trajectory and time evolution of velocity of a steel sphere colliding onto a glass wall immersed in silicon oil RV10, corresponding to Case $\mathrm{St}_n = 152$ of Table \ref{tbl:gondrparam}.
The model is able to accurately reproduce this case. Moreover, the large discrepancy for the numerical solution in the absence of lubrication model illustrates the importance of including it. Note that at each impact the particle has a lower Stokes number: $\mathrm{St}_{n,\mathrm{1st\,b}} = 152$, $\mathrm{St}_{n,\mathrm{2nd\,b}} = 81$, $\mathrm{St}_{n,\mathrm{3rd\,b}} = 23$ and $\mathrm{St}_{n,\mathrm{4th\,b}} = 10$.

\begin{figure}[htp!]
        \centering
   \subfigure[]{
        \centering
        \includegraphics[width=0.45\columnwidth]{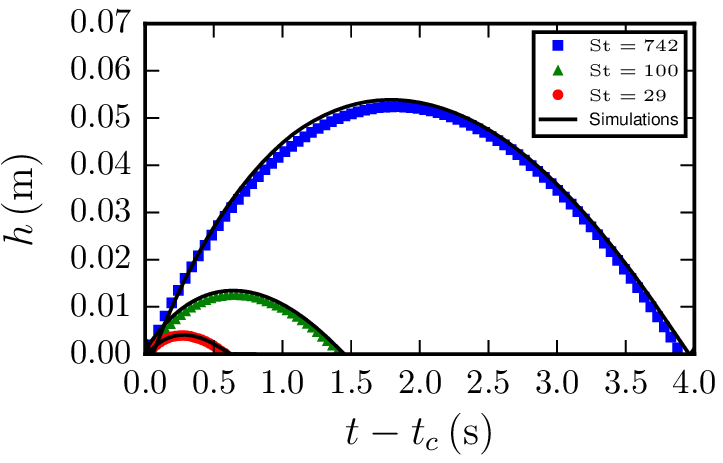}
        \label{fig:bouncing2a}}
\quad
   \subfigure[]{
         \centering
        \includegraphics[width=0.45\columnwidth]{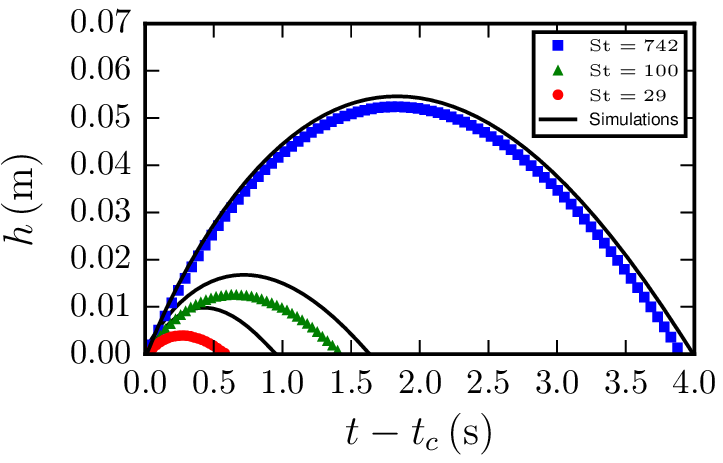}
        \label{fig:bouncing2b}}
   \caption{(Color online) Trajectories obtained from simulations of particles colliding onto a planar surface in silicon oil, for different impact Stokes numbers with (a) and without (b) closure for lubrication interactions. Experimental data from \citet{Gondret2002}.}
   \label{fig:bouncing2}
\end{figure}

Figure \ref{fig:bouncing2a} compares our simulations to the experimental data of \citet{Gondret2002} of the first bounce of steel spheres colliding onto planar surfaces in silicon oil at different impact Stokes numbers. In the cases for which $L_y$ was not sufficiently large for the particle to reach its terminal velocity before colliding with the wall, we imposed an initial velocity to the particle to ease the convergence of the velocity to its terminal value. For extreme cases of a highly inertial $\mathrm{St}_n = 742$ and highly viscous $\mathrm{St}_n = 29$ flow the resolution was increased to $D_p/\Delta x = 32$.

\begin{table}[htp!]
\caption{Properties of the fluids and solid spheres used in the experiment of \citet{Gondret2002} and computational parameters of the numerical simulations.}
\centering
\begin{tabular}{ccccccccccc}
\hline
\hline
Case & $D_p \, [\mathrm{mm}]$ & $\rho_p \, [\mathrm{kg/m^3}]$ & $e_{n,d}$ & $\mu \, [\mathrm{cP}]$ & $\rho_f \,[\mathrm{kg/m^3}]$ & $D_p/\Delta x$ & $\mathrm{Cou}$ & $r_{\Delta t}$ & $N$ \\
\hline
$\mathrm{St}_n = 742$  & $5$ & $7800$ & $0.97$ & $5  $ & $920$ & $32$ & $0.5$ & $50$ & $8$ \\
$\mathrm{St}_n = 152$  & $3$ & $7800$ & $0.97$ & $10 $ & $935$ & $16$ & $0.2$ & $50$ & $8$ \\
$\mathrm{St}_n = 100$  & $4$ & $7800$ & $0.97$ & $20 $ & $953$ & $16$ & $0.2$ & $50$ & $8$ \\
$\mathrm{St}_n = 29 $  & $6$ & $7800$ & $0.97$ & $100$ & $965$ & $32$ & $0.5$ & $50$ & $8$ \\
\hline
\hline
\end{tabular}
\label{tbl:gondrparam}
\end{table}

As expected and shown in Figure \ref{fig:bouncing2b}, the deviation from the experimental data for the simulations without lubrication closure is more significant for smaller Stokes numbers due to the increasing importance of viscous effects. The simulations show a good agreement with the experimental data for this wide range of Stokes numbers.

\subsubsection*{Sensitivity of the results to the collision time and, time step and sub-stepping}

In the following we explore the sensitivity of the model to the computational parameters that govern the collision time and temporal integration of the fluid and particle motion. These parameters are prescribed collision time, amount of sub-stepping and time step of the overall numerical algorithm. Let us consider the trajectory of Figure \ref{fig:bouncing_all} as the reference case for this sensitivity analysis, with focus on the first bounce (the subsequent will be influenced by how realistically the first is reproduced). We performed a set of simulations with parameters shown in Table \ref{tbl:sensitivity}.

\begin{table}[htp!]
\caption{Computational parameters used for the sensitivity study.}
\centering
\begin{tabular}{c c c c c l}
\hline
\hline
Case & $\mathrm{Cou}$ & $r_{\Delta t}$ & $N$ & $\delta_{ij,n}^{max}/{\Delta x} \, (\%)$ & Notes  \\
\hline
{REF} & $0.2$   & $50$  & $8$   & $33.6$ & Reference case                                     \\
{SA1} & $0.6$   & $50$  & $5$   & $63.0$ & Larger $\Delta t$, smaller $N$                     \\
{SA2} & $0.2$   & $1$   & $8$   & $33.6$ & No sub-stepping                                    \\
{SA3} & $0.2$   & $50$  & $1$   & $4.16$ & Small $N$                                          \\
{SA4} & $0.025$ & $50$  & $8$   & $4.16$ & Smallest $\Delta t$, same $T_n$ as SA3             \\
{SA5} & $0.1$   & $50$  & $16$  & $33.6$ & $\Delta t$ between REF and SA5, same $T_n$ as REF  \\
{SA6} & $0.025$ & $50$  & $64$  & $33.6$ & Same $\Delta t$ as SA4, same $T_n$ as REF          \\
\hline
\hline
\end{tabular}
\label{tbl:sensitivity}
\end{table}

\begin{figure}[htp!]
        \centering
   \subfigure{
        \centering
        \includegraphics[width=0.45\columnwidth]{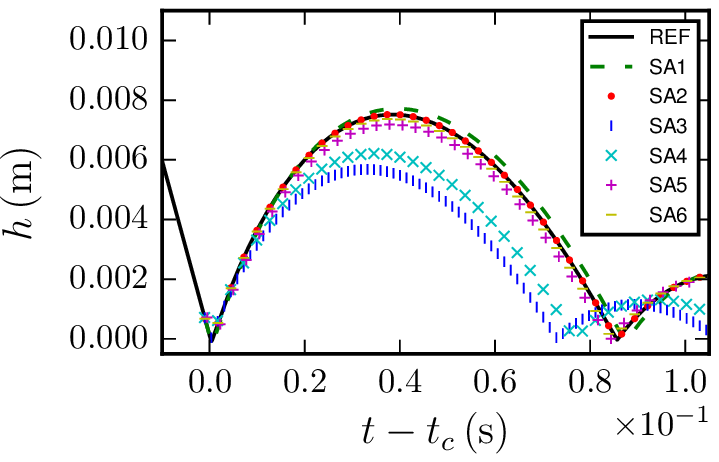}
   \label{fig:sensitivity1}}
\quad
   \subfigure{
        \centering
        \includegraphics[width=0.45\columnwidth]{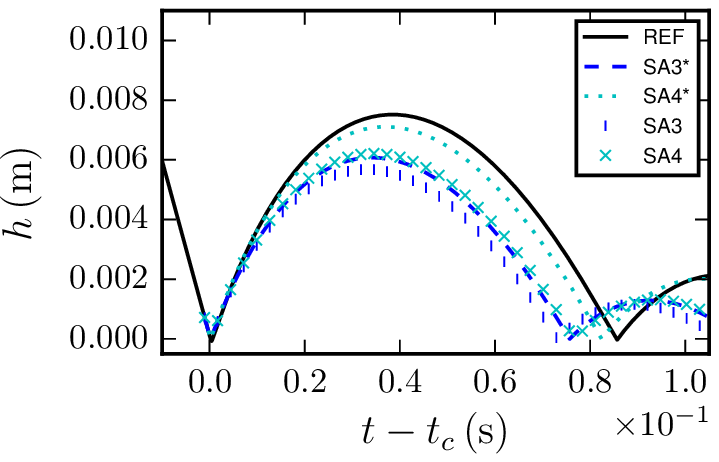}
   \label{fig:sensitivity2}}
   \caption{(Color online) Sensitivity analysis to the time step, sub-stepping and stretching of the collision time (a) and outcome of cases {SA3} and {SA4} when problematic Lagrangian forcing points are excluded from the IBM forcing scheme (b). Computational parameters in Table \ref{tbl:sensitivity}.}
\end{figure}

Figure \ref{fig:sensitivity1} presents the outcome of this set of simulations. The trajectory corresponding to case {SA1} compares well with the one of {REF}, which shows that a collision which takes $5$ Navier-Stokes can still be realistically reproduced. 

The trajectories of cases {SA2} and {REF} cannot be distinguished; this shows that sub-stepping is not required to better resolve the collision and lubrication force corrections in this case because the time step of the Navier-Stokes solver is sufficiently small. 

In case {SA3} the collision time takes exactly one time step of the Navier-Stokes solver, which has the same value that the one of {REF}. Although the collision force and lubrication corrections are resolved due to the sub-stepping, the trajectory obtained from this simulation differs significantly from the reference case. This is mostly a consequence of an over-estimation of the drag force from the IBM when the surrounding fluid does not adapt itself gradually to the abrupt change in particle velocity due to a collision, as illustrated hereafter. 

Decreasing the time step of {SA3} while keeping the stiffness fixed ({SA4}) allows the fluid to adapt itself to the changes in particle velocity. However, the simulation also over-estimates the drag force acting on the particle. We further show with cases {SA5} and {SA6} that the over-estimation of the drag force is not consequence of an inconsistency problem, because the simulations, for the same particle stiffness, converge monotonically to {SA6} with decreasing time step.

The discrepancy of the solution for the stiff particle of case {SA4} is caused by a loss of conservation properties of the interpolation kernel used by the IBM when its stencil, for a certain Lagrangian forcing point, overlaps with the one of another particle or with a solid wall \cite{Kempe2012IBM}. This issue becomes significant for considerably high particle stiffnesses, where more problematic forcing points continue to perform interpolation/spreading operations in a inconsistent manner throughout the entire collision time. Figure \ref{fig:sensitivity2} shows simulations for cases {SA3*} and {SA4*} with the same parameters as the ones of {SA3} and {SA4}, but excluding from the forcing scheme Lagrangian forcing points with a distance to the wall smaller than $\Delta x$ (procedure similar to what is suggested in \cite{Kempe2012IBM}). Indeed, simulation {SA3*} still yields an over-estimated drag force, whereas {SA4*} yields the realistic bouncing trajectory with a difference in the peak of the trajectory of $2.5\%$ from {REF}. 

This illustrates that the realistic bouncing trajectory can only be obtained if the surrounding fluid is allowed to adapt itself to the changes in particle velocity. Hence, we decided to ensure that the fluid phase adapts itself to the changes in particle velocity by avoiding excessively high values of particle stiffness. Note that for the reference case the maximum overlap is already significantly small, about one third of a grid cell.

\subsection{Wet head-on collisions}

The previous validation gives a fine indication of the realism of the approach used to simulate a wet head-on collision. On contrary, the experimental curves of $e_n/e_{n,d}=f(\mathrm{St})$ - benchmark often used to validate these models - depend on the definition of impact and rebound velocities that are used to compute $e_n$. If, for instance, we define $u_{in,n}$ as the terminal velocity, and $u_{out,n}$ as the maximum velocity after impact, for the case $\mathrm{St}_n=152$ of Table \ref{tbl:gondrparam} we obtain $e_n=0.85$; considerably different from the experimentally measured value of $0.78$. To circumvent this problem one can define impact and rebound velocities which agree with the frame rates used in the measurements \cite{Ardekani2008}. We therefore use the impact velocity and rebound velocities at the instants $t-t_{c} = \mp f^{-1}$, respectively; where $f$ is a frequency related to the temporal resolution of the experiment.

\subsubsection*{Particle-wall collisions}

We simulated particle-wall collisions in a viscous liquid for several values of $\mathrm{St}_n$ and compared the resulting normal coefficients of restitution $e_n$ to the experimental data of \citet{Joseph2001}.

\begin{figure}[htp!]
   \centering
   \includegraphics[width=0.45\columnwidth]{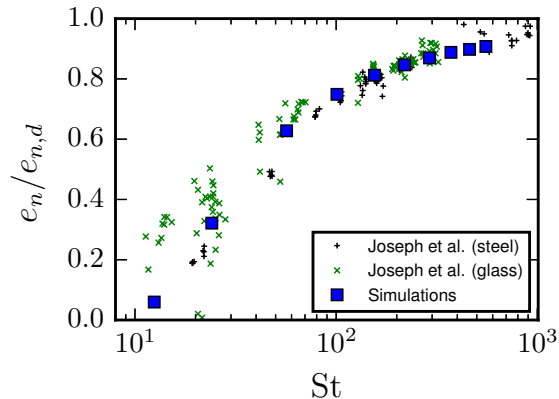}
   \caption{(Color online) Normal, wet coefficients of restitution for particle-wall collisions. The experimental data were normalized with the value $e_{n,d}=0.97$ measured in the reference.}
   \label{fig:normrest}
\end{figure}

The computational domain has dimensions of $L_x/D_p\times L_y/D_p\times L_z/D_p = 12\times 24\times 12$. Similarly to the previous cases, the particles are placed at a distance $y/D_p=L_y-1.5R_p$ and their motion driven by gravity. We simulated steel spheres colliding onto a planar surface in silicon oil RV20 (physical parameters are listed in Table \ref{tbl:gondrparam}). The Stokes number was varied by varying the particle diameters from $1.5 \, \mathrm{mm}$ to $10\,\mathrm{mm}$. We used a value of $f=500\,\mathrm{Hz}$, which complies with the frequency of image acquisition of the experiment. Figure \ref{fig:normrest} shows the results. 

The numerical simulations agree with the experiments for the entire range of impact Stokes numbers. This agreement is expected after the finer validation of the previous section, and careful definition of impact and rebound velocities.

\subsubsection*{Particle-particle collisions}

For inter-particle collisions we reproduced the pendulum experiment of \citet{Yang2006} by colliding a moving projectile particle with a steady target particle. Spheres of the same size and material were centered in a computational box with dimensions $L_x/D_p\times L_y/D_p\times L_z/D_p = 6\times 12\times 6$ and separated in the $y$-direction by a distance of $4 \, D_p$. Similarly to \citet{Simeonov2012}, we force an acceleration $g$ to the projectile particle to mimic the release mechanism of the experiment. 

The physical parameters are comparable to the experiments of head-on collisions of steel spheres in aqueous solutions of glycerol: $e_{n,d}=0.97$, $\rho_p = 7780 \, \mathrm{kg/m^2}$, $D_p=12.7\,\mathrm{mm}$, $\mu = 45\, \mathrm{cP}$, $\rho_f = 1125 \, \mathrm{kg/m^2}$. \citet{Yang2006} defined the rebound and impact velocities at instants corresponding to a value of $f=100\,\mathrm{Hz}$. 

The \textit{binary} impact Stokes number, defined as $\mathrm{St}_{ij,n}\equiv (1/9)\rho_p u_{ij,n} D_p/\mu$ for two equal spheres of the same material, was changed by varying the projectile particle's acceleration from $g = 0.02\times 9.81$ to $10\times 9.81 \, \mathrm{m/s^2}$. We used a value of $\varepsilon_{\sigma}=10^{-5}$ to resolve the lubrication interaction in the thin gap-width between these smooth particles. This value agrees with the order magnitude of the size of the asperities (${\cal O}(0.1)-{\cal O}(0.01)\,\mathrm{\mu m}$ \cite{Yang2006}). These small values together with the fact that the target particle is freely mobile (numerical solution more sensitive to errors when compared to a collision with a wall or a fixed particle) make this benchmark a valuable test for the overall methodology.  Resolving the lubrication layer of the interacting particles at such a small scale required a time step dictated by $\mathrm{Cou}=0.1$ for a resolution of $D_p/\Delta x=16$, and a sub-stepping ratio of $r_{\Delta t} = 1000$. For values of $\mathrm{St}_n$ higher than ${\cal O}(100)$, the resolution required to describe the dynamics of the intervening film is higher. Hence $D_p/\Delta x$ was increased to $32$, with a time step dictated by $\mathrm{Cou}=0.5$.

\begin{figure}[htp!]
        \centering
   \subfigure{
        \centering
        \includegraphics[width=0.45\columnwidth]{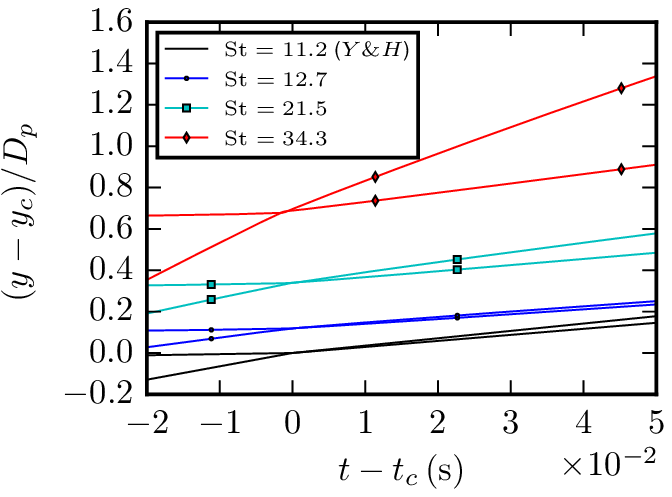}
        \label{fig:pp_separation}}
\quad
   \subfigure{
         \centering
        \includegraphics[width=0.45\columnwidth]{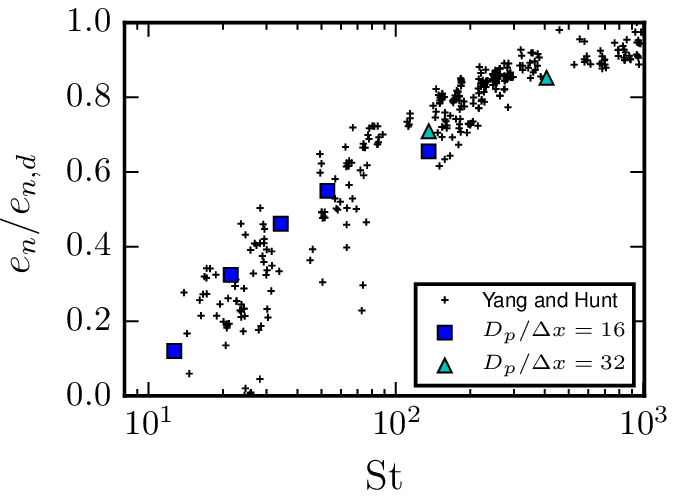}
        \label{fig:pp_en}}
   \caption{(Color online) Trajectories of the particles' contact points (results of the numerical simulations were shifted vertically for clarity). The solid line was extracted from \cite{Yang2006} (a). Wet coefficients of restitution for particle-particle collisions (b). The experimental data were normalized with the value $e_{n,d}=0.97$ measured in the reference.}
   \label{fig:normrest_pp}
\end{figure}

Figure \ref{fig:pp_separation} presents the trajectories of the particles' contact points (results of the numerical simulations shifted vertically for clarity). For very small impact Stokes numbers, the momentum transferred to the target particle is not sufficiently high for it to overtake the viscous drag and travel independently. \citet{Yang2006} observed that this is the case for $\mathrm{St}_{ij,n} \lesssim 10$, where the particles tend to move as a pair with constant separation distance. This is shown in Figure \ref{fig:pp_separation} for cases $\mathrm{St}=11.8$ (measured experimentally) and $\mathrm{St}=12.7$ obtained from a numerical simulation. The good agreement between the numerical simulation and the experiment is a strong indicator of the success of the overall method to resemble this viscous limit. In particular, it gives a finer assessment of the realism of the lubrication closure. Furthermore, the simulations with values of binary impact Stokes number considerably larger than $10$, $\mathrm{St}=21.5$ and $\mathrm{St}=34.3$ do not show this trend, which is consistent with the experimental observations.

Finally, Figure \ref{fig:normrest_pp} compares the computed effective binary coefficient of restitution from the numerical simulations to the experiments. The necessity of increasing the spatial resolution of the simulation for a binary impact Stokes number of $\mathrm{St}=135$ is also illustrated by showing the outcome of this case with both resolutions. Increasing the resolution becomes more important in this case than in particle-wall interactions due to the requirement of an accurate description of the interacting dynamics of the two particles through short-range hydrodynamic interactions.

The agreement with the experimental data further supports the validity of our approach. We should note that extra computational overhead ($\mathrm{Cou}=0.1$ for $D_p/\Delta x = 16$) was required for reproducing these results, when compared to particle-wall collisions.

\subsection{Oblique collisions}

\label{subsec:oblq_results}

Finally, we validated our model for oblique particle-wall collisions in a dry system and in viscous liquids. We use the experimental data of \citet{Joseph2004} of oblique particle-wall collisions in air and aqueous solutions of glycerol. The collisional properties parameters of the particles agree with their experiments and are described together with the other physical parameters of the simulations in Table \ref{tbl:obliqueparam}. The computational domain and particle's initial position is the same of the previous simulations of particle-wall collisions. The particle motion is driven by an imposed acceleration with direction $\mathbf{e}_g = -\sin(\phi_{in})\mathbf{e}_y-\cos(\phi_{in})\mathbf{e}_z$, to yield the desired incidence angle. The magnitude of the particle acceleration was set to $g = 10\times 9.81 \, \mathrm{m/s^2}$ to ensure that the glass spheres collide with an impact Stokes number of ${\cal O}(1000)$, comparable to the experimentally measured values. The results for immersed collisions of steel spheres show little sensitivity to the choice of the value of the acceleration due to the small value of the coefficient of sliding friction.

\begin{table}[htp!]
\caption{Physical and computational parameters for the simulations of oblique particle-wall collisions.}
\centering
\begin{tabular}{ccccccccc}
\hline
\hline
Material & $D_p$ & $e_{n,d}$ & $e_{t,d}$ & $\mu_c$ & $\mu_{c,wet}$ & $\rho_p$ & $\rho_f$ & $\mu$ \\
\hline
steel & $2.5\, \mathrm{mm}$& $0.97$ & $0.34$ & $0.11$ & $0.02$  & $7800\, \mathrm{kg/m^3}$ & $998\, \mathrm{kg/m^3}$ & $1\, \mathrm{cP}$\\
glass & $2.5\, \mathrm{mm}$& $0.97$ & $0.39$ & $0.10$ & $0.15$  & $2540\, \mathrm{kg/m^3}$ & $998\, \mathrm{kg/m^3}$ & $1\, \mathrm{cP}$\\
\hline
\hline
\end{tabular}
\label{tbl:obliqueparam}
\end{table}

\begin{figure}[htp!]
        \centering
   \subfigure{
        \centering
        \includegraphics[width=0.45\columnwidth]{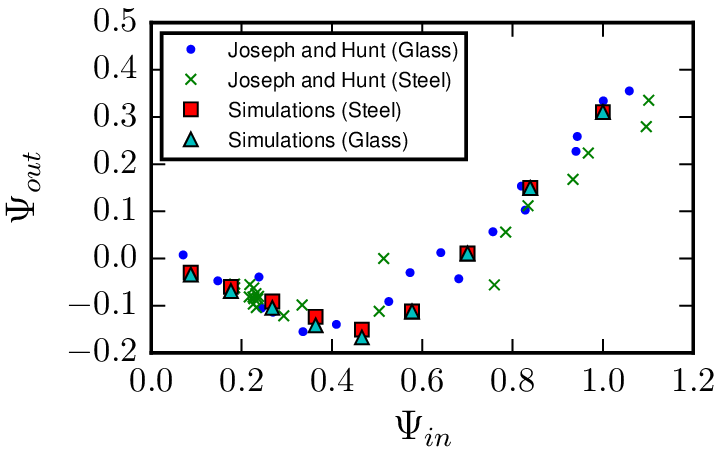}
        \label{fig:oblique_joseph_dry}}
\quad
   \subfigure{
         \centering
        \includegraphics[width=0.45\columnwidth]{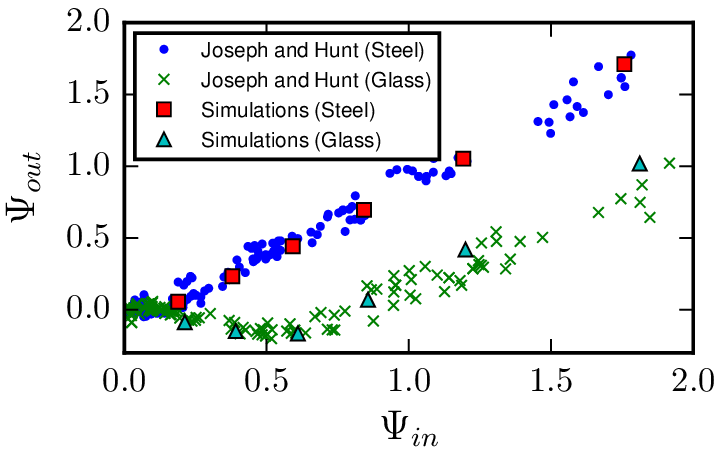}
        \label{fig:oblique_joseph_wet}}
   \caption{(Color online) Results of for oblique collision simulations in a dry system (a), and in a viscous liquids (b). Experimental data of \citet{Joseph2004}.}
   \label{fig:oblique_joseph}
\end{figure}

Figure \ref{fig:oblique_joseph} shows a comparison between the normalized incidence and rebound angles obtained from oblique collisions between steel and glass spheres. 

The simulations agree well with the experimental data for the entire range of incidence angles. This is an expected consequence of the fact that the model uses the macroscopic properties of these collisions as input parameters.

\section{Conclusions and outlook}
\label{sec:conclu}
We presented and validated a collision model for fully-resolved 4-way coupled simulations of flows laden with finite-size solid particles. There are three types of particle-particle or particle-wall interactions that must be reproduced in such simulations: (1) long-range hydrodynamic interactions; (2) short-range hydrodynamic interactions; and (3) solid-solid contact.

The long-range hydrodynamic interactions are computed by a Navier-Stokes solver where we used an IBM for an efficient representation for the particles. Other approaches that require a closure for small inter-particle/particle-wall distances (e.g., Lagrangian-multiplier or Lattice-Boltzmann methods) could have also been used.

Short-range hydrodynamic interactions are also partly resolved by the IBM. However, the discrete nature of these numerical methods together with the necessity of a computationally efficient implementation typically require a closure model for lubrication interactions. For the cases addressed here, the only lubrication interaction that requires modeling is the squeezing of fluid through the thin gap between two approaching particles or a particle approaching a wall. To achieve this we used a two parameter model: for normalized gap-widths smaller than a value $\varepsilon_{\Delta x}$ we introduce a correction based on asymptotic expansions of analytical solutions of particle-particle/-wall interactions in the Stokes regime. This value is obtained by determining the gap-width for which our numerical method is unable to reproduce the lubrication interaction. The second parameter, $\varepsilon_{\sigma}$, accounts for roughness effects for even smaller gap-widths.

Finally, solid-solid contact is modeled through a linear soft-sphere collision model capable of stretching the collision time, to avoid computational overhead in the calculation of the collision force. The model constants are analytically related to the three input parameters of the model described by \citet{Walton1993}, which are widely reported in the literature. The model can be extended to accommodate more complex mechanics such as adhesion or plasticity for the normal force, or static and dynamic friction for the tangential force. However, these features are in general not required in 4-way coupled simulations of flows with finite-size particles at small/moderate solid volume fractions.

We validated our methodology against several benchmark experiments and the results show a good quantitative agreement. The simulations of the bouncing trajectory of a spherical particle colliding onto a planar surface \cite{Gondret2002} show that the lubrication force corrections, combined with the collision model are sufficient for reproducing a realistic bouncing velocity. Subsequently, we successfully reproduced experimental data for the normal coefficient of restitution as a function of the impact Stokes numbers for head-on particle-wall \cite{Joseph2001} and particle-particle collisions \cite{Yang2006}. Finally, our simulations of oblique particle-wall collisions in dry and wet systems agree quantitatively with the experimental data of \citet{Joseph2004} for the entire range of incidence angles. We reserve further validations of the overall model for flows with many particles where both lubrication and friction play an important role for a future publication.

The physical realism and computational efficiency of our method allows for massive fully-resolved simulations of particle-laden flows with 4-way coupling.

\section{Acknowledgments}

We acknowledge Gustavo Joseph \cite{Joseph2001}, \cite{Joseph2004}, Philippe Gondret \cite{Gondret2002} and Fu-Ling Yang \cite{Yang2006} for kindly providing their experimental data in electronic format. This work was supported by the Portuguese Foundation for Science and Technology under the grant FRH/BD/85501/2012.

\bibliography{bibfile}

\end{document}